\documentclass[aip,jcp,reprint]{revtex4-1}
\usepackage{amsmath}
\usepackage{afthd}
\usepackage{mathptmx}
\usepackage{textcomp,dcolumn,xspace}
\usepackage{graphicx}
\usepackage[ansinew]{inputenc}
\usepackage{hyperref}
\hypersetup{bookmarksnumbered=true,bookmarksopen,bookmarksopenlevel=1,
            colorlinks,linkcolor=blue,citecolor=blue,urlcolor=black,
            pdfauthor={V. Holten and M. A. Anisimov},
			bookmarksdepth=2,pdfpagemode=UseOutlines,pdftitle={Entropy-driven liquid-liquid separation in supercooled water},
            pdfprintscaling=None}

\newcommand{\ea}{\textit{et al.}\xspace}
\newcommand{\figref}[1]{Fig.~\ref{#1}}

\renewcommand*{\eqref}[1]{equation~(\ref{#1})}

\newcommand{\di}{\mathrm{d}}

\newcommand*{\ts}[1]{_\text{#1}}
\newcommand*{\tu}[1]{^\text{#1}}
\newcommand*{\kb}{k\ts{B}}
\newcommand*{\Dt}{\Delta \hat{T}}

\newcommand*{\Dp}{\Delta \hat{P}}

\newcommand*{\Pc}{P_\text{c}}
\newcommand*{\Tc}{T_\text{c}}

\newcommand*{\rhoc}{\rho_\text{c}}

\newcommand*{\xe}{x\ts{e}}

\newcommand*{\mA}{\mu\tu{A}}
\newcommand*{\mB}{\mu\tu{B}}
\newcommand*{\GnA}{G\tu{A}}
\newcommand*{\GnB}{G\tu{B}}
\newcommand*{\GnBA}{G\tu{BA}}

\newcommand*{\mBA}{\mu\tu{BA}}

\begin{document}

\title{Entropy-driven liquid--liquid separation in supercooled water}
\affiliation{Institute for Physical Science \& Technology and Department of Chemical \&
Biomolecular Engineering, University of Maryland, College Park, Maryland
20742, USA}
\author{V. Holten}
\author{M. A. Anisimov}
\email[e-mail: ]{anisimov@umd.edu}
\date{26 February 2013}

\begin{abstract}
Twenty years ago Poole et al. (Nature \textbf{360}, 324, 1992) suggested that the
anomalous properties of supercooled water may be caused by a critical point that
terminates a line of liquid--liquid separation of lower-density and higher-density water.
Here we present an explicit thermodynamic model based on this hypothesis, which describes
all available experimental data for supercooled water with better quality and with fewer
adjustable parameters than any other model suggested so far. Liquid water at low
temperatures is viewed as an `athermal solution' of two molecular structures with
different entropies and densities. Alternatively to popular models for water, in which
the liquid--liquid separation is driven by energy, the phase separation in the athermal
two-state water is driven by entropy upon increasing the pressure, while the critical
temperature is defined by the `reaction' equilibrium constant. In particular, the model
predicts the location of density maxima at the locus of a near-constant fraction (about
0.12) of the lower-density structure.
\end{abstract}

\maketitle

\noindent Cold and supercooled water have been the subject of intensive experimental,
theoretical and computational studies for the last several decades. Still, the famous
statement of 1972 by Franks `of all known liquids, water is probably the most studied and
least understood'\cite{franks1972} remains topical. This is especially true for
metastable supercooled water, which is now a focal point of debates. On the other hand,
there is a growing interest in the prediction of properties of supercooled water. In
particular, in applied atmospheric science it is commonly accepted that the uncertainties
in numerical weather prediction and climate models are mainly caused by poor
understanding of properties of water in tropospheric and stratospheric clouds, where
liquid water can exist in a deeply supercooled state\cite{rosenfeld2000,heymsfield1993}.
Reliable prediction of properties of supercooled water is also important for
cryobiology\cite{song2010}.

A provocative, but thermodynamically consistent, view on the global phase behaviour of
supercooled water was expressed in 1992 by Poole \ea\cite{poole1992}. According to this
view, deeply in the supercooled region, just below the line of homogeneous ice
nucleation, there could exist a critical point of liquid--liquid coexistence (LLCP) that
would terminate the line of first-order transitions between two liquid aqueous phases,
low-density liquid (LDL) and high-density liquid (HDL) (Fig.~\ref%
{fig:phasediagram}). The anomalies in the heat capacity, the compressibility, and the
thermal-expansion coefficient experimentally observed upon supercooling%
\cite{angell1982,kanno1979,kanno1980,tombari1999,arc00,hare86,hare87,mishima2010,mishima2010}
thus might be associated with this critical point, even if it is inaccessible.

\begin{figure}
\includegraphics{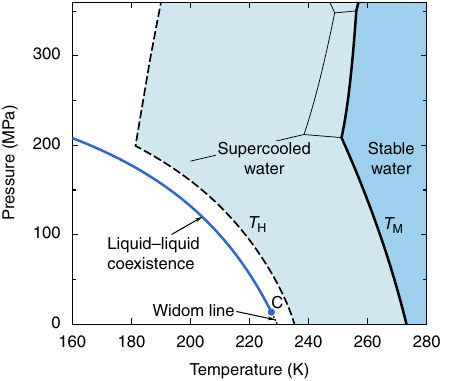}
\caption{\label{fig:phasediagram}\textbf{Hypothetical phase diagram of cold water.}
Supercooled water exists between the melting
temperature\cite{wagner2011} $T\ts{M}$
and the homogeneous ice nucleation temperature\cite{kanno2006} $T\ts{H}$.
Below $T\ts{H}$, there may exist
a liquid--liquid critical point, marked by `C', which terminates a liquid--liquid coexistence curve.
The location of this critical point is shown as predicted by our model.
The adopted location of the liquid--liquid coexistence curve is
close to similar suggestions of Mishima\cite{mishima2000,mishima2010}
and Kanno and Angell\cite{kanno1979}.
The extension of the coexistence curve into the one-phase region is the line of
maximum fluctuations of the order parameter, the Widom line\cite{xu2005}. Thin solid lines represent phase
boundaries between the different ices\cite{bridgman1912,kell1968}.}
\end{figure}

Intriguing liquid--liquid phase separation and the existence of the second critical point
in water still remain a plausible hypothesis which needs further verification. In view of
the inaccessibility of the LLCP to direct experiments, development of an equation of
state, based on a solid physical concept and able to accurately describe all available
experimental data, might help in resolving the supercooled-water dilemma.

In this paper, we offer an approach to thermodynamics of phase separation in supercooled
water alternative to common views. Following Bertrand and Anisimov\cite{bertrand2011}, we
assume that liquid water is a non-ideal athermal `solution' of two supramolecular
arrangements, which undergoes phase separation driven by non-ideal entropy upon increase
of the pressure. In the athermal two-state model, the non-ideality driven by entropy
determines the critical pressure, whereas the critical temperature follows from the
condition of `chemical reaction' equilibrium. We have developed an explicit equation of
state, which is based on the athermal two-state model and which describes all available
experimental data for supercooled water, both H$_{2}$O and D$_{2}$O, with better quality
and with fewer adjustable parameters than any other model suggested so far. We have also
shown that the popular two-state `regular-solution' scenario, in which phase separation
is energy-driven, fails to describe the experimental data on water.

\section{How a pure liquid can unmix}
The only possible fluid--fluid phase transition in simple one-component substances, such
as argon or methane, is separation into liquid and gas. This transition is driven\ by
attraction forces between molecules. The critical temperature of the liquid--gas
separation is uniquely proportional to the depth of the intermolecular interaction
potential, which is a superposition of the attraction and repulsion potentials. The
simplest model of the liquid--gas transition is the lattice gas, which, despite its
simplicity, reflects the most important features of fluid phase behaviour
\cite{leeyang52}. A central concept in physics of phase transitions is the notion of the
order parameter\cite{landaulifshitzstatistical}. The order parameter in fluids is the
difference between the densities of liquid and vapour. Thus, in simple one-component
fluids only one liquid phase and only one critical point are possible. To observe a
liquid--liquid separation one needs a mixture of two or more species. Remarkably, the
same lattice-gas model that describes the liquid--gas transition in pure substances can
also explain liquid--liquid demixing in weakly-compressible binary solutions. In binary
liquid solutions the order parameter is the difference in concentrations of two liquid
phases, while the critical temperature is proportional to the difference in the depths of
the interaction energies between like and unlike molecules.

However, such a simple fluid-phase behaviour is not a law of nature. As explained by
Mishima and Stanley\cite{mishima1998review}, if the intermolecular potential of a pure
fluid could exhibit two minima, the interplay between these minima may define the
critical temperature $\Tc$ and pressure $\Pc$ of liquid--liquid separation. Application
of this picture to water implies that the second, liquid--liquid, critical point is
driven by molecular interaction energy just like the liquid--gas critical point.

The hypothesized existence of two liquid states in pure water can be globally viewed in
the context of polyamorphism, a phenomenon that has been experimentally observed or
theoretically suggested in silicon, liquid phosphorus, triphenyl phosphate, and in some
other molecular-network-forming substances \cite{mcmillan2004,mcmillan2006}. Commonly,
polyamorphism in such systems is described as energy-driven. However, there is an
ambiguity in terminology adopted in Refs.~\onlinecite{mcmillan2004} and
\onlinecite{mcmillan2006}, where the term `density, entropy-driven' is used for an
energy-driven phase separation. There is an example of entropy-driven liquid
polyamorphism, isotropic Blue Phase III in chiral liquid crystals\cite{anisimov1998}.

The thermodynamic relation between the molar volume change $\Delta V$ and the latent heat
(enthalpy change) of phase transition, $Q =T\Delta S$ (where $\Delta S$\ is the molar
entropy change) is given by the Clapeyron equation $\di P/\di T=Q/T\Delta V=\Delta
S/\Delta V$. Therefore, the relation between the volume/density ($\rho =1/V)$ change and
the latent heat/entropy change is controlled by the slope of the transition line in the
$P$--$T$ plane.

Two features make the second critical point in water phenomenologically different from
the well-known gas--liquid critical point. The negative slope of the liquid--liquid phase
transition line in the $P$--$T$ plane in supercooled water (Fig.~\ref{fig:phasediagram})
means, in accordance with the Clapeyron equation, that the higher-density liquid water is
the phase with larger entropy. A large value of this slope at the critical point (about
30 times greater than that for the gas--liquid transition at the critical point)
indicates the significance of the entropy change with respect to the density change, and,
correspondingly, the importance of the entropy fluctuations. Secondly, supercooled water
tends to separate upon pressurizing. The relative significance of the entropy change,
combined with a high degree of cooperativity of hydrogen bonds\cite{stokely2010}, suggest
that the liquid--liquid phase separation in water near the LLCP may be driven by entropy
rather than by energy.

The Clapeyron equation itself does not provide an answer whether the liquid--liquid
transition in pure substances is energy-driven or entropy-driven. To answer this question
one should examine the source of non-ideality in the free energy. Bertrand and
Anisimov\cite{bertrand2011} introduced the concept of a `lattice liquid', as opposed to a
lattice gas, an imaginary one-component liquid which exhibits liquid--liquid separation
upon pressurizing with a vertical liquid--liquid transition line in the $P$--$T$ plane,
thus without the density change. The order parameter in a lattice liquid is the entropy
and the `ordering field', conjugate to the order parameter, is the temperature. The phase
transition in a lattice liquid is purely entropy-driven. The critical temperature (the
same as the temperature of the liquid--liquid transition line) is defined by
thermodynamic equilibrium as zero ordering field, but not by the interaction potential.
In real water, as seen in Fig. 1, the slope of the hypothesized LDL--HDL transition line
in supercooled water changes from very steep at higher temperatures to relatively flat at
lower temperatures. It is thus tempting to assume that the liquid--liquid separation in
water may represent a special kind of liquid polyamorphism, intermediate between two
limiting cases: mostly entropy-driven phase separation (lattice liquid-like) near the
critical point and mostly energy-driven (lattice gas-like) separation into two amorphous
states observed in water at about 140 K\cite{loerting2011PCCP}. We show in this paper
that the actual behaviour of supercooled water appears to be much closer to the
lattice-liquid behaviour than to the lattice-gas behaviour.

The entropy-driven separation of a lattice liquid can be further specified in terms of a
two-state model\cite{bertrand2011}. The pure liquid is assumed to be a mixture of two
interconvertible states or structures of the same molecules, whose ratio is controlled by
thermodynamic equilibrium. The existence of two states does not necessarily mean that
they can separate. If these states form an ideal `solution', the liquid will remain
homogeneous at any temperature or pressure. However, if the solution is non-ideal, a
positive excess Gibbs energy, $G\tu{E} = H\tu{E}-T S\tu{E}$, would cause phase
separation. If the excess Gibbs energy is associated with a heat of mixing $H\tu{E}$, the
separation is energy-driven. If the excess Gibbs energy is associated with an excess
entropy $S\tu{E}$, the separation is entropy-driven. The entropy-driven nature of such a
separation means that the two states would allow more possible statistical
configurations, and thus higher entropy, if they are unmixed.

Two-state models for liquid water have a long history, dating back to the 19th
century\cite{whiting1884,roentgen}. More recently, two-state models have become popular
to explain liquid polyamorphism \cite{mcmillan2004,mcmillan2006,vedamuthu1994}.
Ponyatovsky \ea\cite{ponyatovsky1998} and Moynihan\cite{moynihan1997} assumed that water
could be considered as a `regular binary solution' of two states, which implies that the
phase separation is driven by energy. None of these two-state models for water has been
used for a quantitative description of available experimental data.

\section{Virtual liquid--liquid criticality in supercooled water}
There is no direct experimental evidence of the LLCP in real water, but it is indirectly
supported by thermodynamic arguments based on density measurements\cite{mishima2010} and
by critical-like anomalies of thermodynamic response
functions\cite{angell1982,kanno1979,kanno1980}. The known existence of two states of
glassy water\cite {loerting2011PCCP}, as well as experiments on nano-confined water,
which does not crystallize\cite{zhang2011,nagoe2010}, are also consistent with the
possibility of a `virtual' liquid--liquid separation. This possibility is also supported
by simulations of some water-like models, such as ST2\cite{sciortino2011}. The exact
location of the liquid--liquid critical point in these models is
uncertain\cite{stokely2010}. Moreover, for the mW model\cite{molinero2009} it has been
recently shown that spontaneous crystallization occurs before a possible liquid--liquid
separation could equilibrate\cite{limmer2011,moore2011}. However, the anomalies observed
in the metastable region of the mW
model\cite{molinero2009,moore2009,moore2011,limmer2011} might still be associated with
the existence of a virtual LLCP in the unstable region.

The first attempt to develop an equation of state for supercooled water, based on the
assumption that the LLCP exists, and on the asymptotic theory of critical
phenomena\cite{fisher1983,behnejad2010}, was made by Fuentevilla and
Anisimov\cite{fuentevilla2006} and further elaborated and clarified by Bertrand and
Anisimov\cite{bertrand2011}. In particular, both works estimated the LLCP critical
pressure below 30~MPa, much lower than most of simulated water-like models predicted.
Holten \ea\cite{holtenSCW} used the same asymptotic equation of state, also in a
mean-field approximation\cite{holtenMF2012}, but introduced the noncritical backgrounds
of thermodynamic properties in a thermodynamically consistent way. The resulting
correlation represents all available experimental data for supercooled water, H$_{2}$O
and D$_{2}$O, within experimental accuracy, thus establishing a benchmark for further
developments in this field. However, there is a concern regarding the application of the
asymptotic theory to a broad range of temperatures and pressures including the region far
away from the assumed critical point. Such an extension makes the description of
experimental data inevitably semi-empirical since all non-asymptotic physical features
are absorbed by the adjustable backgrounds of thermodynamic properties. This fact
underlines the need to develop a closed-form theoretically-based equation of state which
would satisfy the asymptotic critical anomalies and, at the same time, describe regular
behaviour far away from the critical region.

\section{Water as an athermal solution of two states}
We assume pure liquid water to be a mixture of two interconvertible states or structures
A (HDL) and B (LDL). The fraction of molecules in state B is denoted by $x$, and is
controlled by the `reaction'
\begin{equation}\label{eq:reaction}
    \text{A}\rightleftharpoons \text{B}.
\end{equation}
The states A and B could correspond to different arrangements of the
hydrogen-bonded network.\cite{eisenbergbook_mixturemodels} The Gibbs energy
per molecule $G$ is the sum of the contributions from both states,
\begin{equation*}
    G=(1-x)\mA+x\mB=\mA+x\mBA.
\end{equation*}
where $\mA$ and $\mB$ are the chemical potentials of A and B in the mixture. The variable
$x$ is conjugate to $\mBA\equiv\mB-\mA$. If A and B form an athermal non-ideal solution,
the Gibbs energy of the mixture is
\begin{equation}\label{eq:gibbs}
    \frac{G}{\kb T}=\frac{\GnA}{\kb T}+x\frac{\GnBA}{\kb T}+x\ln x+(1-x)\ln(1-x)+\omega x(1-x),
\end{equation}
and the chemical-potential difference is then
\begin{equation*}
    \mBA=\GnBA+\kb T\left[ \ln \frac{x}{1-x}+\omega (1-2x)\right] ,
\end{equation*}
where $\GnBA\equiv \GnB-\GnA$ is the difference in Gibbs energy per molecule between pure
configurations A and B, $\kb$ is Boltzmann's constant, and $\omega =\omega(P)$ is the
interaction parameter, which depends on pressure but not on temperature.

Considering $x$, the fraction of B, as the `reaction coordinate' or `extent of reaction'
\cite{prigogine1954}, the condition of chemical reaction equilibrium,
\begin{equation*}
    \left( \frac{\partial G}{\partial {x}}\right) _{T,P}=\mBA=0,
\end{equation*}
yields the equilibrium constant $K=K(T,P)$ of reaction (\ref{eq:reaction}) as
\begin{equation*}
    \ln K=\frac{\GnBA}{\kb T}.
\end{equation*}
and defines the equilibrium fraction $\xe$ through
\begin{equation}\label{eq:xe}
    \ln K+\ln \frac{\xe}{1-\xe}+\omega (1-2\xe)=0.
\end{equation}
The expression (\ref{eq:gibbs}) combined with the equilibrium condition $x=\xe$ is our
equation of state for the two-state supercooled water. The non-ideality of the two-state
mixture is entirely associated with the excess entropy of mixing $S\tu{E}=-\omega
x(1-x)$, while the heat (enthalpy) of mixing is zero. The parameter $\omega $ determines
the critical pressure through its pressure dependence, which we approximate as
\begin{equation*}
    \omega =2+\omega_0 \Dp,
\end{equation*}
where $\Dp = (P-\Pc)/\rhoc\kb\Tc$ with a subscript `c' denoting critical parameters. An
alternative, regular-solution model would have a factor $w/\kb T$ in place of $\omega $
in \eqref{eq:gibbs}, where the interaction parameter $w$ would determine the critical
temperature. For a regular solution, the excess entropy is zero, while the heat of mixing
is $H\tu{E} = w x(1-x)$, thus making phase separation purely energy driven.

The conditions for the critical point of liquid--liquid equilibrium,
\begin{equation*}
    \biggl(\frac{\partial ^{2}G}{\partial \xe^{2}}\biggr)_{T,P}=0,\qquad
    \biggl(\frac{\partial ^{3}G}{\partial \xe^{3}}\biggr)_{T,P}=0,
\end{equation*}
yield in the case of the athermal-solution model the critical parameters $x\ts{c}=1/2$
and $\omega (\Pc)=2$. At pressures below the critical pressure, $\omega <2$ and the Gibbs
energy versus fraction $x=\xe$ shows a single minimum. Above the critical pressure,
$\omega >2$ and there are one or two minima in the Gibbs energy, depending on the value
of the equilibrium constant $K(T,P)$. If there are two minima, the minimum with the
lowest Gibbs energy represents stable equilibrium, and the other minimum corresponds to a
metastable state. When $\ln K(T,P) =0$, both minima have the same Gibbs energy,
representing two-phase equilibrium. The critical temperature $\Tc$ and the location of
the liquid--liquid transition (LLT) curve in the $P$--$T$ plane are thus determined by
the dependence of the equilibrium constant $K$ on temperature and pressure. Since the LLT
curve is defined as an analytical function of temperature and pressure $\ln K(T,P)=0$,
this function is to be obtained from the experimental data. In Supplementary Section~1 we
explain how we match the LLT curve of the two-state model to the experimentally expected
shape and location. The resulting expression for $\ln K$ is
\begin{equation*}
    \ln K=\lambda(\Dt+a\Dp+b\Dt\Dp),
\end{equation*}
where $\Dt = (T-\Tc)/\Tc$. The parameter $a = -\rhoc\kb\di T/\di P$ is the slope of the
LLT curve at the critical point, and $b$ determines the curvature. The parameter
$\lambda$ is proportional to the heat of reaction (\ref{eq:reaction}), while the product
$\upsilon = \lambda a$ is proportional to the volume change of the reaction. Since
$\lambda/\upsilon = \Delta S/\rhoc\kb\Delta V = Q/T\Delta V$, the heat of reaction is
asymptotically related to the latent heat of phase separation through $\lambda\Delta
x=Q/\kb\Tc$, while $\Delta x = \Delta S/\kb\lambda = Q/\lambda\kb\Tc$ serves as the order
parameter along the LLT.

The Gibbs energy $\GnA$ of the pure structure A defines the background properties and is
approximated as
\begin{equation}\label{eq:background}
    \GnA = \sum_{m,n} c_{mn}(\Dt)^m (\Dp)^n,
\end{equation}
where $m$ (0 to 3) and $n$ (0 to 5) are integers and $c_{mn}$ are adjustable
coefficients. Expressions for thermodynamic properties which follow from our model are
given in Supplementary Section~2.

\section{Effects of critical fluctuations}
Thermodynamics predicts the divergence of fluctuations at the critical point. Entropy
fluctuations are proportional to the isobaric heat capacity $C_P$, volume/density
fluctuations are proportional to the isothermal compressibility $\kappa_T$, and cross
entropy--volume fluctuations are proportional to the thermal expansion coefficient
$\alpha_P$ \cite{landaulifshitzstatistical}. The two-state model described above is
essentially mean-field, not affected by fluctuations. Being expanded near the critical
point in powers of $x-x\ts{c}$, \eqref{eq:gibbs} takes the form of a Landau
expansion\cite{landaulifshitzstatistical}. The procedure to include the effects of
critical fluctuations is well developed and known as crossover theory\cite{behnejad2010}.
This procedure is fully described in Supplementary Section~3. The variables $x-x\ts{c}$
and $P - \Pc$ are renormalized such that the behaviour close to the critical point agrees
with the asymptotic critical behaviour\cite{holtenSCW}, and crosses over to mean-field
behaviour given by the equation of state, (\ref{eq:gibbs}) and (\ref{eq:xe}), far away
from the critical point.

\section{Description of experimental data and discussion}
\begin{figure}
\includegraphics[width=8cm]{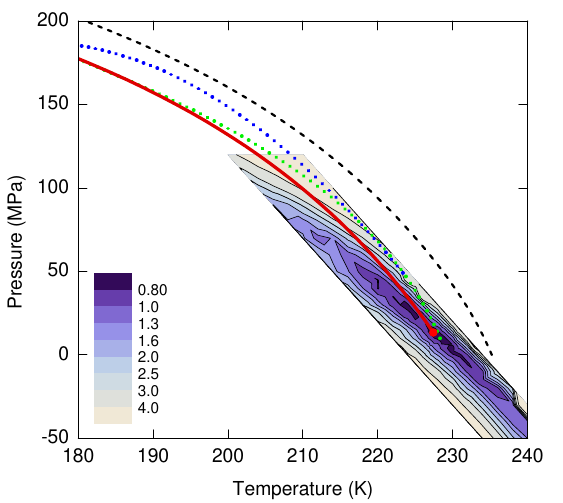}
\caption{\label{fig:contours}\textbf{Optimization of the LLCP location.}
The coloured map shows the reduced sum of squared residuals. The solid red line is the
hypothesized LLT curve. The dashed curve shows the temperature of homogeneous ice nucleation\cite{kanno2006}.
The blue dotted curve is the LLT suggestion by Mishima\cite{mishima2010} and the green dotted curve
is the `singularity' line suggested by Kanno and Angell\cite{kanno1979}.}
\end{figure}

\begin{figure}
\includegraphics{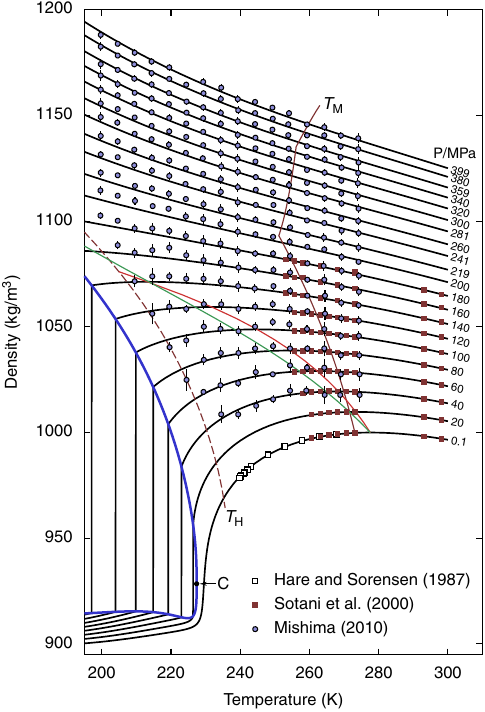}
\caption{\label{fig:density}\textbf{Density of cold and supercooled water as a function of temperature and pressure.}
Black curves are the predictions of the crossover two-state model.
$T\ts{M}$ (thin black) indicates the melting temperature and $T\ts{H}$ indicates the homogeneous nucleation
temperature. The thin blue line is the liquid--liquid
equilibrium curve, with the critical point C.
The red line is the line of maximum density, and the green line is the line of a constant LDL
fraction of about 0.12.
Symbols represent experimental data\cite{mishima2010,sotani2000,hare87}. Mishima's
data\cite{mishima2010} have been shifted by at most 0.3\% to bring them into agreement
with data for stable water, as described in Ref.~\onlinecite{holtenSCW}.}
\end{figure}

\begin{figure*}
\includegraphics{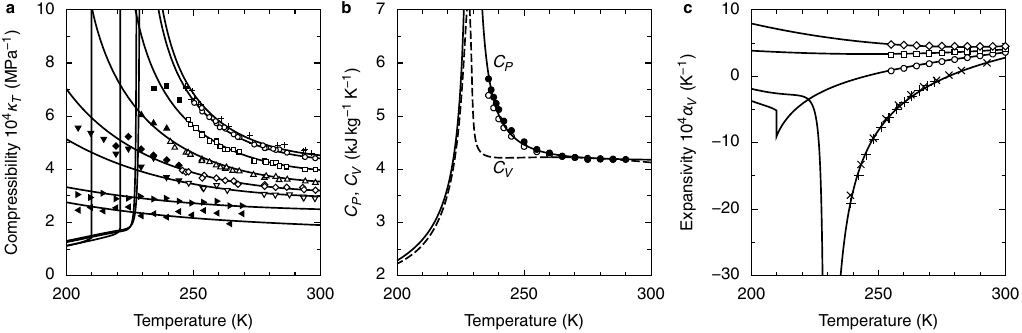}
\caption{\textbf{Response functions as a function of temperature.} \textbf{a}, Isothermal
compressibility\cite{speedy1976,kanno1979,mishima2010}. Pressures, from top to bottom:
0.1~MPa, 10~MPa, 50~MPa, 100~MPa, 150~MPa, 190~MPa, 300~MPa, 400~MPa. \textbf{b}, Heat
capacity at constant pressure $C_P$ (open circles: data from Archer and
Carter\cite{arc00}, closed circles: data from Angell \ea\cite{angell1982}) and at
constant volume $C_V$ (calculated) at 0.1~MPa. See Supplementary Section~5 for more
details. \textbf{c}, Thermal expansivity\cite{hare86,hare87,terminassian1981}. Pressures,
from top to bottom: 380~MPa, 200~MPa, 100~MPa, 0.1~MPa. In \textbf{a, b,} and \textbf{c,}
the curves are the prediction of the crossover two-state model, and the symbols represent
experimental data.} \label{fig:responsefunctions}
\end{figure*}

We have fitted both the mean-field and the crossover formulation of our two-state model
to experimental data, as described in Supplementary Section~4. We adopt the location of
the liquid--liquid coexistence parallel to the homogeneous nucleation curve, which
appears to be close to a suggestion of Mishima\cite{mishima2000,mishima2010} based on the
shape of metastable melting curves of different ices, and a `singularity line' of Kanno
and Angell\cite{kanno1979} based on the extrapolation of the compressibility anomalies.
The optimum locations of the critical point form a narrow band in the $P$--$T$ diagram.
The best fit for the critical point is obtained at about 227~K and 13~MPa, with
$\lambda=2.3$ and $\omega_0=0.35$. The LLT curve was chosen to intersect the band of LLCP
locations at the optimal value of the critical pressure (\figref{fig:contours}) where it
has a dimensionless slope of $1/a = 15.3$. This location of the critical point is about
10\% of $\rhoc\kb\Tc$ higher than that optimized by the mean-field approximation (see
Supplementary Section~4A). This shift is induced by critical fluctuations, as follows
from crossover theory.

\begin{figure}
\includegraphics{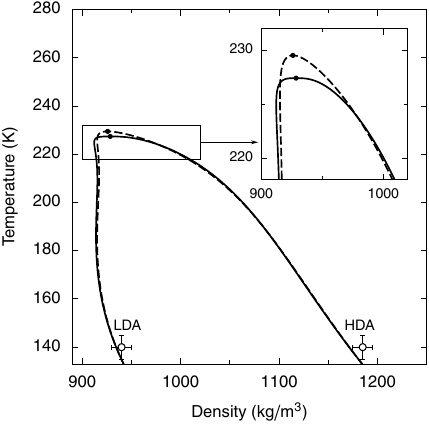}
\caption{\label{fig:coexistence}\textbf{Density along the liquid--liquid transition curve.}
The dashed line represents the mean-field two-state model;
the solid line represents the crossover two-state model.
The open circles are the densities of the
low-density amorphous (LDA) and high-density amorphous (HDA) phases of water at 200~MPa\cite{loerting2011PCCP}.
One can notice that the crossover LLT curve is flatter than the LLT curve in the mean-field approximation
and that the actual position of the critical point is shifted to a lower temperature by critical
fluctuations.}
\end{figure}

The description of the density and the response functions are shown in
Figs.~\ref{fig:density} and~\ref{fig:responsefunctions}. A significant improvement
compared to the previous works\cite{bertrand2011,holtenSCW,holtenMF2012} is that our
equation of state does not show any sign of additional thermodynamic instability beyond
the liquid--liquid separation. We now believe that the mechanical and thermal instability
below the LLT, reported previously\cite{bertrand2011,holtenSCW,holtenMF2012}, were
associated with the asymptotic nature of the equation of state, which caused negative
backgrounds of the compressibility. The number of adjustable background coefficients in
\eqref{eq:background} is now 14, to be compared with 16\cite{holtenSCW,holtenMF2012} (see
Supplementary Table S1). Another improvement is the agreement of the coexistence
densities, shown in \figref{fig:coexistence}, with the experimental densities of
low-density and high-density amorphous water. We also fitted our model to all available
experimental data for supercooled D$_2$O with the same quality as for H$_2$O (see
Supplementary Section~4B). With the same number of adjustable parameters, a
regular-solution two-state model fails (about ten times higher root-mean-square error) to
describe experimental data on supercooled water.

\begin{figure}
\includegraphics{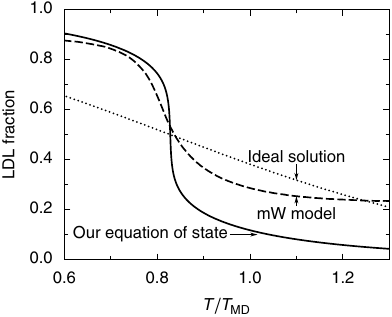}
\caption{\label{fig:fraction}\textbf{Fraction of molecules in the low-density state.}
The fraction $x$ is shown for the two-state model at 0.1~MPa, in the case of an athermal solution (solid) and
an ideal solution (dotted). The dashed curve is the fraction of four-coordinated water molecules,
i.e., the fraction of water molecules with four neighbours, for mW water simulations
performed by Moore and Molinero\cite{moore2009} (see Supplementary Section~6).
The temperature is scaled by the temperature of maximum density $T\ts{MD}$ (250~K for the
mW model and 277~K for real water).}
\end{figure}

In Fig.~\ref{fig:fraction}, we compare the LDL fraction $x$ predicted by our equation of
state and that obtained in simulations of the mW model by Moore and
Molinero\cite{moore2009}. Remarkably, both results show a similar temperature dependence.
In contrast, this fraction for an ideal LDL/HDL solution would be almost a linear
function of temperature. The experimental maximum density line is located approximately
along the line of constant LDL fraction $x=0.12$ as shown in \figref{fig:density}.

We confirm the previous finding\cite{fuentevilla2006,bertrand2011} of a critical pressure
that is much lower than found in simulations\cite{sciortino2011}. We believe that the low
critical pressure reflects the entropy-driven nature of liquid--liquid criticality in
supercooled water. As shown by Stokely \ea\cite{stokely2010}, the LLCP critical pressure
is determined by the ratio of hydrogen-bond cooperativity and hydrogen-bond covalent
strength. The higher the ratio, the lower the critical pressure.

The necessity to develop a microscopic model for water, which would be consistent with
its athermal two-state character, and which would clarify the microscopic nature of the
order parameter, $(x-x\ts{c})/x\ts{c}$, is evident.

Finally, the problem of the influence of the hypothesized liquid--liquid separation on
homogeneous ice nucleation remains unresolved. The fact that the LLT curve is located
just below the homogeneous nucleation curve and imitates its shape suggests that
homogeneous ice nucleation may be caused by the entropy/structure fluctuations associated
with the liquid--liquid transition. A connection between the change in the structure of
liquid water and the crystallization rate of ice was shown for the mW
model\cite{moore2011}. Whether the LLCP of simulated water-like models is in the
metastable region or just projected to be in the unstable region, the liquid--liquid
criticality still may be responsible for the observed anomalies in the accessible domain.

\section*{References}
\bibliography{vincentnew,supercooled}

\section*{Acknowledgments}
We acknowledge collaboration with C.E. Bertrand, D.A. Fuentevilla, J.~Kalová, J.~Leys,
and J.V. Sengers. We thank C.A. Angell, P.G. Debenedetti, O. Hellmuth, O. Mishima, V.
Molinero, and H.E. Stanley for useful discussions. The research has been supported by the
Division of Chemistry of the US National Science Foundation under Grant No. CHE-1012052.
The research of V. Holten was also supported by the International Association for the
Properties of Water and Steam.

\end{document}